\def\year{2019}\relax
\documentclass[letterpaper]{article} 
\usepackage{aaai19}  
\usepackage{times}  
\usepackage{helvet} 
\usepackage{courier}  
\usepackage[hyphens]{url}  
\usepackage{graphicx} 
\usepackage{graphicx}  
\usepackage{subcaption}
\usepackage{boldline}
\usepackage{soul}
\usepackage{amsmath}
\usepackage{amssymb}
\usepackage{mathtools}
\usepackage{multirow}
\usepackage[mathscr]{euscript}

\begin{document}
%
\title{Hierarchical Context enabled Recurrent Neural Network for Recommendation}
\author{	
	Kyungwoo Song$^{1}$\thanks{Equal contribution.}, Mingi Ji$^{1}$\footnotemark[1], Sungrae Park$^{2}$ \and Il-Chul Moon$^{1}$ \\
    $^{1}$ 	Korea Advanced Institute of Science and Technology (KAIST), Korea \\
    $^{2}$ Clova AI Research, NAVER Corp., Korea  \\
	\{gtshs2,qwertgfdcvb\}@kaist.ac.kr, sungrae.park@navercorp.com, icmoon@kaist.ac.kr \\}
\maketitle
\begin{abstract}
A long user history inevitably reflects the transitions of personal interests over time. The analyses on the user history require the robust sequential model to anticipate the transitions and the decays of user interests. The user history is often modeled by various RNN structures, but the RNN structures in the recommendation system still suffer from the long-term dependency and the interest drifts. To resolve these challenges, we suggest HCRNN with three hierarchical contexts of the global, the local, and the temporary interests. This structure is designed to withhold the global long-term interest of users, to reflect the local sub-sequence interests, and to attend the temporary interests of each transition. Besides, we propose a hierarchical context-based gate structure to incorporate our \textit{interest drift assumption}. As we suggest a new RNN structure, we support HCRNN with a complementary \textit{bi-channel attention} structure to utilize hierarchical context. We experimented the suggested structure on the sequential recommendation tasks with CiteULike, MovieLens, and LastFM, and our model showed the best performances in the sequential recommendations.
\end{abstract}

\section{Introduction}
\begin{figure}[t!]
    \includegraphics[width=1.0\columnwidth]{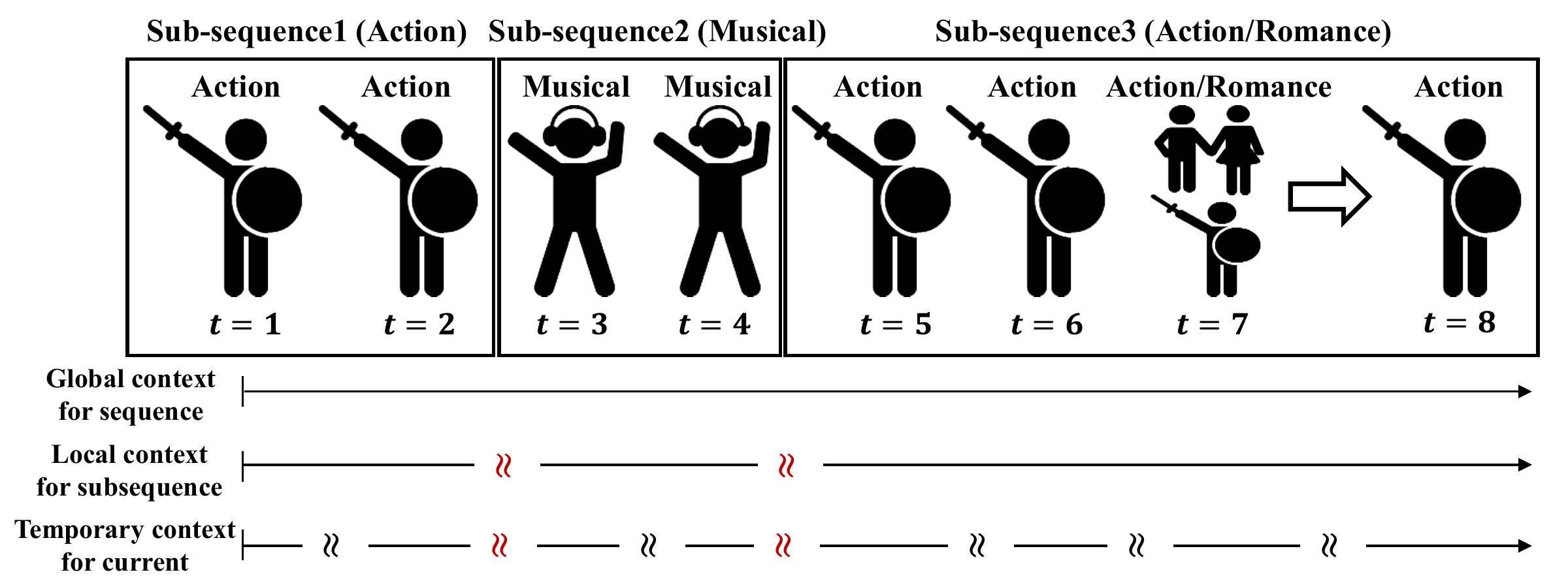}
    \caption{The long user history contains multiple hierarchical context; global context, local context, and temporary context. To take into account the user's interest drift, the temporary context must change at every point (black wave) but should change more when the new sub-sequence starts (red wave). The wave means the change of interest, and red wave means a more drastic change than the black wave.
        Moreover, the interest drift should be considered in the hierarchical context. For example, in the figure above, we can see that the user's primary interest is the action movie, given the global and local context.
        Therefore, even if a movie, whose genre is action and romance, comes out at $t = 7$, we can recommend an action movie at $t=8$ rather than a romance movie if we consider the hierarchical context.}
    \label{fig:intro_example}
\end{figure}

A user \textit{history} is a sequence of user orders or clicks, and the history represents the user's interest. Given this user history, many services such as movie recommendations, music streaming services, etc., are interested in recommending the next most likely click item. When we perform this recommendation, it has been assumed that the user's interest can be hierarchically ranging from general interest to a temporary, specific need as shown in Figure \ref{fig:intro_example}. Here, these hierarchical interest dynamics are defined as 1) the global context for the entire sequence; 2) the local context for a sub-sequence, such as a click-stream of a site visit with a few or dozens of clicks; 3) and the temporary context for a transition of items. The assumption on the hierarchical contexts has been partially reflected in NARM that models the attention of the general interest \cite{li2017neural}; and STAMP that directly predicts the next item by considering temporary contexts \cite{liu2018stamp}.

The recently proposed models with recurrent neural network (RNN) structures have focused on modeling the local context of sub-sequences \cite{wu2017recurrent,smirnova2017contextual,yin2016adapting}. For example, GRU4REC \cite{hidasi2015session} utilized a gated recurrent unit (GRU) \cite{Cho2014} with a ranking based loss, to emphasize the best item selection. This model started modeling the interest dynamics with general structure, GRU, but the general structure can be further modified to model the hierarchical interest dynamics. Another example is NARM whose attention mechanism is one way of modeling the user's global context. This attention mechanism emphasizes a specific previous record to consider for the next recommendation, and this mechanism enables the long sequence modeling. However, this mechanism could be better if we include modeling on a sudden interest drift of users. In contrast to GRU4REC and NARM, STAMP is optimized to model the short interest drift of users without any recurrent structures \cite{liu2018stamp}. STAMP embeds only the right-before item for temporary interest and the cumulative summary of previous items for general (or global) interest with two feed-forward networks. 
The STAMP model can be further improved if the structure takes into account the hierarchical interaction between global interest and temporary interest as Figure \ref{fig:intro_example}.

This paper proposes Hierarchical Context enabled RNN (HCRNN) which models the hierarchical interest dynamics within a modified RNN structure. To our knowledge, this is the first proposal to operate the hierarchical contexts of interest dynamics with a modified RNN cell structure that optimizes both keeping the global/local context and accepting the temporary drift. 
HCRNN is similar to the LSTM's mechanism of modeling long-term and short-term memory, separately; but there are inherent differences, as well. 

HCRNN does not generate the temporary context from either global or local context. LSTM uses the cell state to produce its corresponding hidden state that is a short-term memory, and with this structure, the hidden state tends to be a subset of cell state. However, if we assume the global interest dynamics can be fundamentally different from the temporary transition, i.e., an sudden purchase order out of consistent long purchase history, we need to separate the long-term memory and the short-term memory. 

HCRNN independently maintains the local context and the temporary context, and they interact each other only in the gate (Eq. 15, 17, 18) and attention (Eq. 13) while LSTM does not independently keep the short-term hidden output. 
For hierarchical context modeling, the global and local contexts need to contain more abstract information than the temporary context.
For this purpose, we proposed a new structure to generate the local context that combines the advantages of topic modeling and memory network \cite{sukhbaatar2015end,lau2017topically}.

\begin{figure}[ht]
	\centering
	\includegraphics[width=1.0\columnwidth]{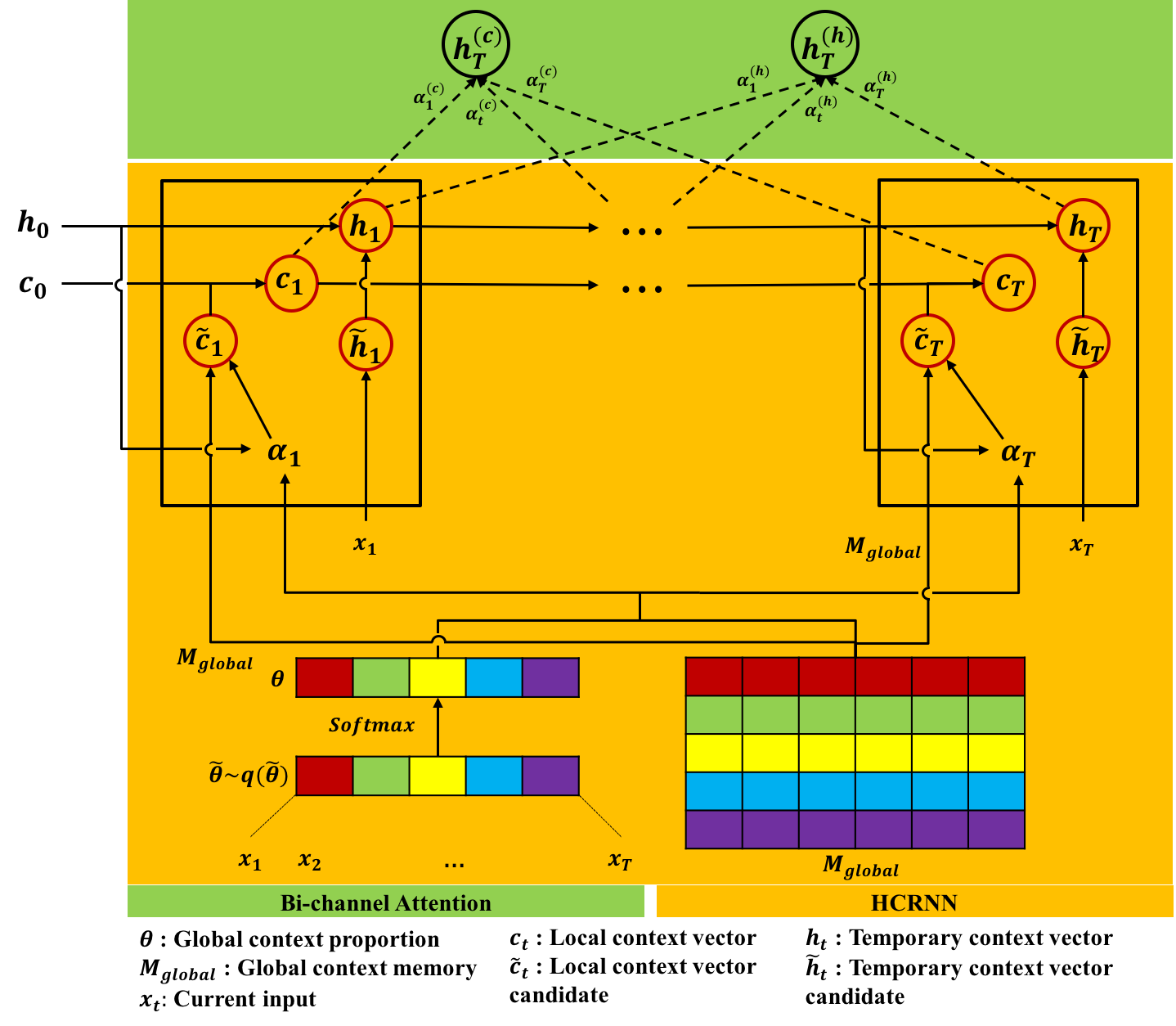}
	\caption{Overall HCRNN and Bi-channel attention structures. Each different color boxes of the $\theta$ is the proportion of $k$-th global context vector, $\theta^{(k)}$. Each row of the $M_{global}$ is the $k$-th global context vector, $M_{global}^{(k)}$. Same color of box in $\theta $ and row in $M_{global}^{(k)}$ mean the same global context.}
	\label{fig:overall_structure}
\end{figure}

As shown in Figure \ref{fig:intro_example}, it is easy to capture the interest drift of the user with a hierarchical context.
In other words, we defined the \textit{interest drift assumption} as ``if the user's local context (for sub-sequence) and the current item are very different, the user's temporary interest drift occurs." We proposed a new gate structure to incorporate this assumption effectively. 
As we propose a modified RNN cell and its outputs with different semantics, we also suggest a modified attention mechanism that is complementary to the proposed cell structure. As the global context becomes a static context, the dynamic context becomes the local and the temporary contexts. Therefore, the attention mechanism will be bi-channel with the local and the temporary contexts, so we named it as the \textit{bi-channel attention}. 
By the combination of the HCRNN cells, the attention weight with the local context is concentrated on the recent history, and the attention weight with temporary context is distributed to the relatively far history.
We have presented an overall structure of HCRNN and bi-channel attention in Figure \ref{fig:overall_structure}. 

\section{Preliminary}
\subsection{Cell Structure of Recurrent Neural Networks} 
\subsubsection{LSTM}
LSTM is a de facto standard of RNNs by enabling the learning from the long-term dependency. A variant of LSTM, or LSTM with peephole connection \cite{Gers2002}, is a typical LSTM structure with emphasis on the modified gating mechanism by accepting the input from the cell state, and LSTM with peephole are used in previous studies \cite{Zhu2017,Neil2016}. The below is the specifications of LSTM with peephole with formulas.
\begin{flalign}
i_{t} &= \sigma_{i}(x_{t} W_{xi}+h_{t-1}W_{hi} + c_{t-1} \odot w_{ci}  +b_{i}) \\
f_{t} &= \sigma_{f}(x_{t} W_{xf}+h_{t-1}W_{hf} + c_{t-1} \odot w_{cf}  +b_{f}) \\
\widetilde{c}_{t} &= x_{t}W_{xc}+h_{t-1}W_{hc}+b_{c} \\
c_{t} &= f_{t} \odot c_{t-1} + i_{t} \odot \sigma_{c}(\widetilde{c}_{t}) \label{eq:LSTSM_ct}\\
o_{t} &= \sigma_{o}(x_{t} W_{xo}+h_{t-1}W_{ho} + c_{t} \odot w_{co}  +b_{o}) \\
h_{t} &= o_{t} \odot \sigma_{h}(c_{t}) \label{eq:LSTM_ht}
\end{flalign}

Here, it should be noted that Eq. \ref{eq:LSTM_ht} generates the hidden variable of LSTM, which we consider a temporary context in HCRNN. Eq. \ref{eq:LSTM_ht} does not have any component of $h_{t-1}$ and it means the high dependency of $h_{t}$ on $c_{t}$. This treatment of connection is hard to consider the semantically different context between the local and the temporary context at the same time.

HCRNN modifies the LSTM structure to treat the generation of the local and the temporary context separately to consider the hierarchical contexts at the same time. Besides, we modified the gate structure to consider the interaction between the hierarchical contexts.

\subsubsection{GRU}
GRU is a simplified version of LSTM with fewer parameters while GRU still supports learning from the long-term dependency. GRU replace cell state $(c_{t})$ and hidden state $(h_{t})$ in LSTM with one hidden state ${(h_{t})}$. The below is the specification of GRU.

\begin{flalign}
z_{t} &= \sigma_{z}(x_{t} W_{xz}+h_{t-1}W_{hz} + b_{z}) \label{eq:GRU_zt} \\
r_{t} &= \sigma_{r}(x_{t} W_{xr}+h_{t-1}W_{hr} + b_{r}) \label{eq:GRU_rt} \\
\widetilde{h}_{t} &= (r_{t} \odot h_{t-1})W_{hh} + x_{t}W_{xh} + b_{h} \\
h_{t} &= (1-z_{t}) \odot h_{t-1} + z_{t} \odot \sigma_{h}(\widetilde{h}_{t})
\end{flalign}

The enabler of GRU mechanism is Eq. \ref{eq:GRU_zt} and \ref{eq:GRU_rt}. Hence, when we seek a new gating mechanism to separate the generation of contexts, we were motivated by adopting such condensed gating mechanisms because HCRNN will inevitably increase the number of trained parameters.

\subsection{Attention on Recurrent Neural Networks}
An RNN representing a context up to the present with a fixed length vector suffers from long-term dependency considerations. For that reason, \cite{Bahdanau2014} proposed an attention mechanism to retrieve the information needed at present among the past information. The RNN attention mechanism is usually based only on the hidden state of an encoder ($h$) and decoder ($s$) of the RNN. $\alpha_{ij} = exp(e_{ij})/\sum_{k=1}^{T}exp(e_{ik})$ is the attention weight at a point $j$ in time $i$, and $\alpha_{ij}$ is determined by $e_{ij} = v_{a}^{T}\sigma(W_{a}s_{i-1}+U_{a}h_{j})$.
We were motivated by adopting such a hidden state, $h$, based attention algorithm because consideration of long-term dependency is important for the sequential recommendation.
Besides, in recommendation tasks where user interest drifts frequently occur, it is also important to consider the recent user history.
For this reason, we have also adopted the local context, $c$, based attention to account for the recent history in the sub-sequence.

\begin{figure*}[ht]
	\centering
	\begin{subfigure}[t]{0.24\textwidth}
		\centering
		\includegraphics[width=\textwidth]{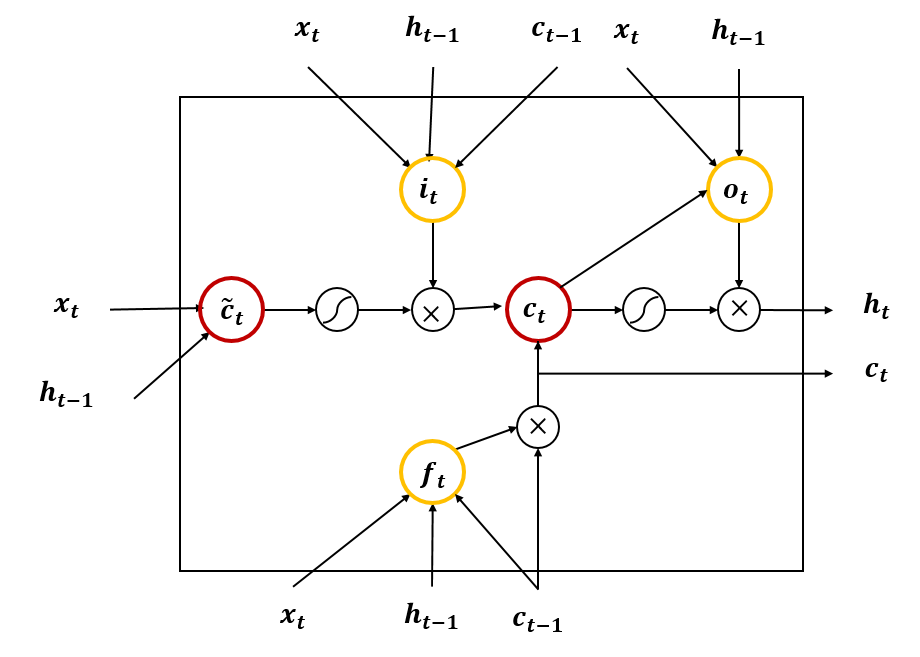}
		\caption{LSTM with peephole}
		\label{fig:LSTM_structure}
	\end{subfigure}
	\begin{subfigure}[t]{0.24\textwidth}
		\centering
		\includegraphics[width=\textwidth]{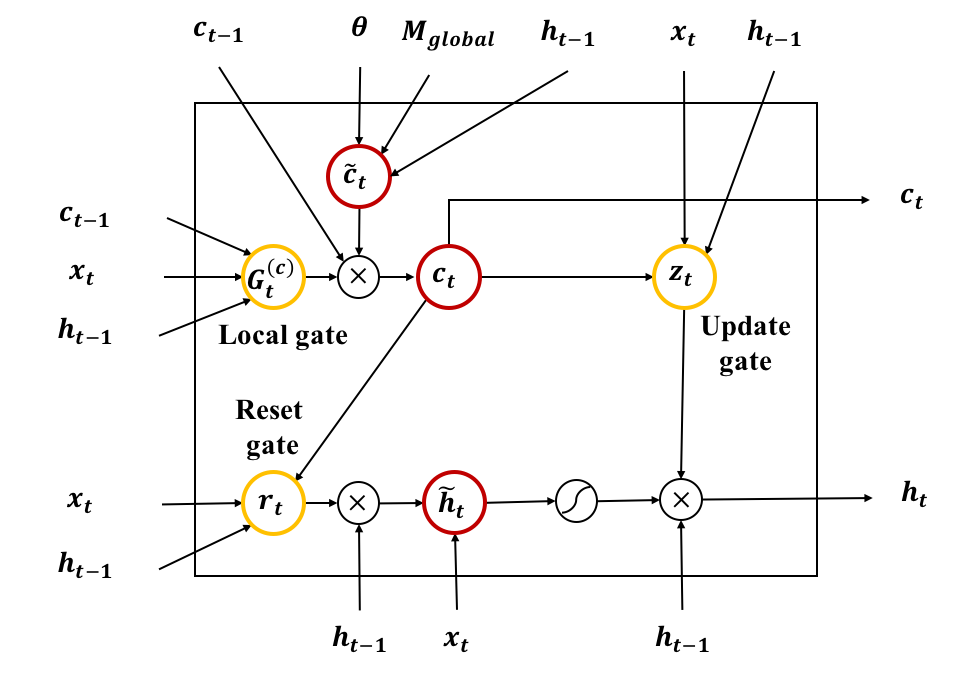}
		\caption{HCRNN-1}
		\label{fig:HCRNN_v1_structure}
	\end{subfigure}
	\hfill 
	\begin{subfigure}[t]{0.24\textwidth}
		\centering
		\includegraphics[width=\textwidth]{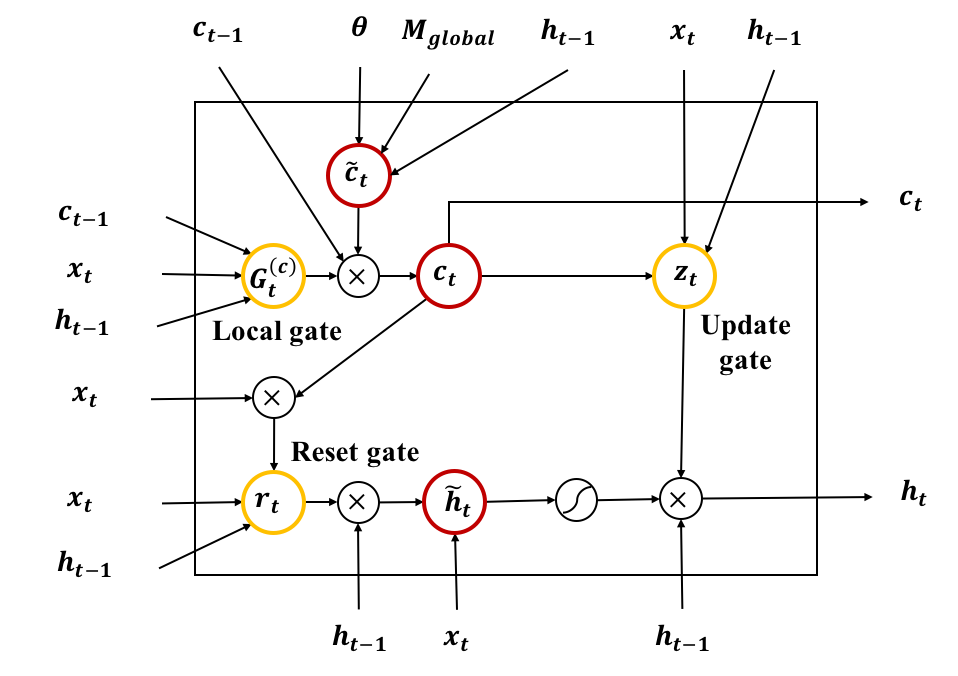}
		\caption{HCRNN-2}
		\label{fig:HCRNN_v2_structure}
	\end{subfigure}
	\hfill 
	\begin{subfigure}[t]{0.24\textwidth}
		\centering
		\includegraphics[width=\textwidth]{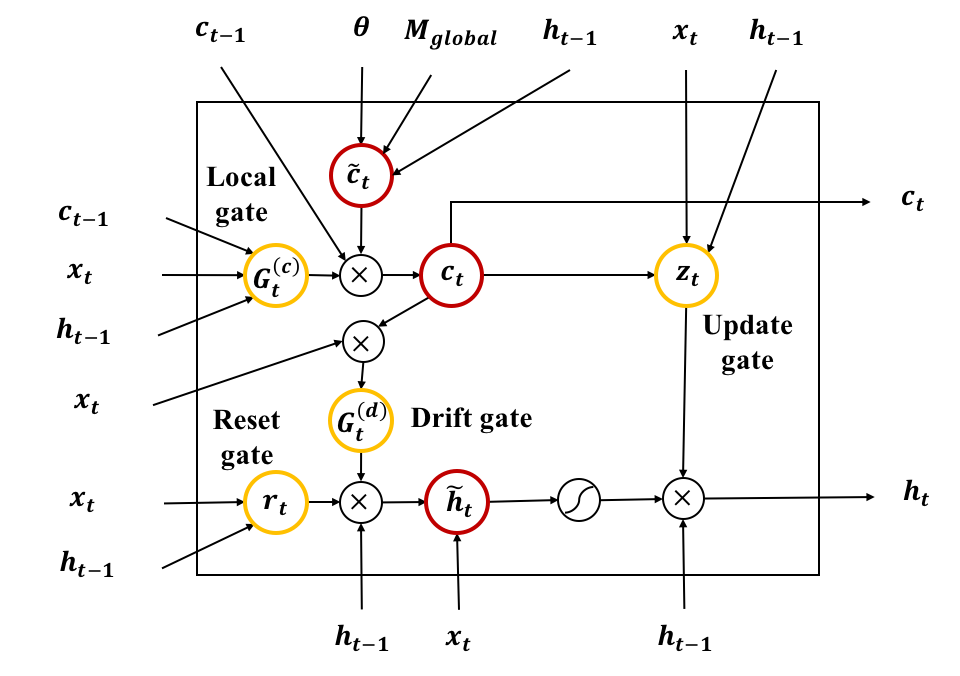}
		\caption{HCRNN-3}
		\label{fig:HCRNN_v3_structure}
	\end{subfigure}
	\caption{HCRNN structure and LSTM with peephole structure. Unlike LSTM, the creation of temporary contexts, $h_{t}$, in HCRNN is separated from local contexts, $c_{t}$. Also, the and the local contexts, $c_{t}$, are designed to be influenced by the item embedding, $x_{t}$, only through the gate or attention structure. For this reason, the local context can have a more abstract context than a temporary context. Besides, we propose a new gate structure to incorporate the interest drift assumption relatively strongly to HCRNN-1, HCRNN-2, and HCRNN-3.}
	\label{fig:LSTM_HCRNN}
\end{figure*}

\section{Methodology}
This paper introduces HCRNN-1, HCRNN-2, HCRNN-3, and bi-channel attention. First, we will explain the overall structure of HCRNN, followed by a detailed modeling of HCRNN and bi-channel attention in order.
\subsection{Hierarchical Context Recurrent Neural Network}
We propose HCRNN, a modification of the RNN structure, to model three hierarchical contexts optimized for recommendations, which we describe in this section. 
\subsubsection{Overall Structure}
We summarize the overall structure of HCRNN cell at three points.
First, $h_{t}$ of LSTM, which corresponds to the temporary context in HCRNN, is generated by the $c_t$, which corresponds to the local context in HCRNN. This generation in LSTM indicates that the temporary context is directly influenced by the current cell state, $c_t$, while HCRNN has no such direct influence to the temporary context, as discussed in \textit{introduction} section. Hence, the generation of the temporary context in HCRNN is detached from the local context, and the local context and temporary context has the connection through the gating mechanism. This makes the creation of temporary contexts more flexible and captures instantaneous interest drift.
\begin{center}
\begin{table}[t]
    {\small
        \hfill{}
        \begin{tabular}{|l|l|}
            \hline
            \textbf{Notation} & \textbf{Description} \\
            \hline
            $|K|$ & Dimension of global context proportion \\
            \hline
            $|D|$ & Dimension of item embedding \\
            \hline
            $|H|$ & The number of hidden units in HCRNN \\
            \hline
            $|I|$ & The number of items \\
            \hline
            $|T|$ & The length of the sequence  \\
            \hline
            $x_t$ & The $t$-th input embedding \\
            \hline
            $\tilde{c}_t$ & The $t$-th local context candidate \\
            \hline
            $c_t$ & The $t$-th local context\\
            \hline
            $\tilde{h}_t$ & The $t$-th temporary context candidate \\
            \hline
            $h_t$ & The $t$-th temporary context \\
            \hline
            $G_t^{(c)}$ & The $t$-th local context gate \\
            \hline
            $G_t^{(d)}$ & The $t$-th drift gate \\
            \hline
            $M_{global}$ & Global context memory\\
            \hline
            $M_{global}^{(k)}$ & The $k$-th global context vector  \\
            \hline
            $\theta$ & Global context proportion \\
            \hline
            $\theta^{(k)}$ & The $k$-th global context proportion \\
            \hline
            $r_t$ & The $t$-th reset gate of HCRNN \\
            \hline
            $z_t$ & The $t$-th update gate of HCRNN  \\
            \hline
            $\alpha_t$ & Global memory attention \\
            \hline
            \multirow{ 2}{*}{$\alpha_t^{(c)}$} & Local context attention weight\\
            &in bi-channel attention \\
            \hline
            \multirow{ 2}{*}{$\alpha_t^{(h)}$} & Temoporary context attention \\
            &weight in bi-channel attention \\
            \hline
            \multirow{ 2}{*}{$W_{c\alpha}^{(1)}, W_{c\alpha}^{(2)}$} & Projection matrices for\\
            & local context attention \\
            \hline
            \multirow{ 2}{*}{$W_{h\alpha}^{(1)}, W_{h\alpha}^{(2)}$} & Projection matrices for\\
            & temporary context attention \\
            \hline
            $W_{emb}$ & Item embedding matrix  \\
            \hline
            $W_{B}$ & Weight for bi-linear decoding \\
            \hline
            $\sigma, \sigma_r, \sigma_z, \sigma_l, \sigma_d$ & Sigmoid activation function \\
            \hline
            $\sigma_h$ & $\tanh$ activation function \\
            \hline
    \end{tabular}}
    \caption{The description for the notation in this paper.}
    \label{table:Notation}
\end{table}
\end{center}
Second, we introduce a new static context in the RNN structure as the global context. HCRNN models the global context as two latent variables of $M_{global}$ and $\theta$. 
Global context memory $M_{global}$ contains global context vector $M_{global}^{(k)}$ for each global context component $k$.
$\theta$ is the global context proportion, which means the weight of each global context in the sequence. In other words, $\theta^{(k)}$ is the proportion of activating a certain part of the global context vector, $M^{(k)}_{global}$. 

This second modification can be used to generate the local context which contains abstract information. We designed a new unified algorithm for the local context creation for unifying the memory network, the topic modeling, and the recurrent structure with attention. This algorithm has three advantages. 1) The local context is generated from a global context memory,$M_{global}$, which is a memory network so that it can contain abstract information. 2) The attention used to generate the local context reflects both global context proportion and global context memory. We generate the local context by reflecting the global context proportion at each timestep. For example, if most of the items in a sequence are action movies, the HCRNN is trained to have a high probability in generating a local context associated with the action movie. 3) When the global context is imported into the current local context in Eq. \ref{eq:HCRNN1_alpha},\ref{eq:HCRNN1_c_candiate}, we utilize the temporary context, $h_t$, so a local context, $c_t$ is adapted to the global context influenced by the temporary context. The HCRNN cell in Figure \ref{fig:overall_structure} illustrate this hierarchical context generation of $M_{global}$, $\theta$, $c_t$, and $h_t$.

Third, we designed the gating structure to reflect the interest drift assumption by hierarchical contexts.
Figure \ref{fig:intro_example} illustrates an occurrence of interest drifts when a user selects an item different from the local context.
To reflect this interest drift assumption, we modified the reset gate in HCRNN-2 as shown in Eq. \ref{eq:HCRNN2_reset}.
Furthermore, HCRNN-3 has a drift gate, $G_{t}^{(d)}$ in Eq. \ref{eq:HCRNN3_drift}, with only local context ($c_{t}$) and current item embedding ($x_t$) while the reset gate is influenced by the previous temporary context ($h_{t-1}$), the local context, and the item embedding, jointly. $G_{t}^{(d)}$ emphasizes the reset initiated only by the interest drift. Figure \ref{fig:LSTM_HCRNN} shows the overall structure of LSTM and HCRNN-1,2,3 structure.

\subsubsection{HCRNN-1}
The first version of HCRNN introduces the global contexts and the modified structure from the LSTM cell. First, as we introduced in the previous section, the global context consists of the global context proportion, $\theta$, and global context memory, $M_{global}$. Here, $\theta$ is similar to the topic proportion in general topic models, such as LDA \cite{Blei2003}, and our model is designed by following the TopicRNN \cite{Dieng2016}. However, unlike TopicRNN, we introduce $M_{global}$ designed as a memory network holding the abstract information, represented as an embedding of each topic, or a global context in the recommendation domain. 

The modification consists of two phases. First, to model the local context candidate $\widetilde{c}_{t}$, we modeled the degree, $\alpha_{t}$, to which we should consider for each global context vector, $M_{global}^{(k)}$, at the current time step.
$\alpha_{t}$ is obtained from the attention mechanism, which is different from the bi-channel attention in \textit{Bi-channel Attention and Prediction} section, based on the previous temporary context $h_{t-1}, M_{global}$ and $\theta$ in Eq. \ref{eq:HCRNN1_alpha}. This attention of $\alpha_{t}$ is an attention mechanism within the HCRNN cell, yet the bi-channel attention is attention outside of the HCRNN cell sequence.
Because $\alpha_{t}$ is computed with the temporary context of $h_{t-1}$, the local context candidate, $\widetilde{c}_{t}$, can fluctuate temporarily. To handle this fluctuation, we formulate a local gate, $G_{t}^{(c)}$ in Eq. \ref{eq:HCRNN1_local_gate}, \ref{eq:HCRNN1_c}. The local gate $G_{t}^{(c)}$, helps local context to change more stable, different with temporary context.

The second phase is modeling the temporary context, $h_{t}$, with the current input, $x_t$, and the previous temporary context, $h_{t-1}$. The temporary context does not directly come from the local context of $c_t$ and the global context of $\theta,M_{global}$. However, the reset gate of $r_{t}$ uses the local and the temporary contexts to reset the components of the temporary context. Additionally, the update gate of $z_{t}$ controls the update with the current input, the local and the temporary contexts. This structure allows the temporary context to focus more on the current input, unlike the local context. 
{\small
\begin{flalign}
&\widetilde{\theta} \sim q(\widetilde{\theta}) = \mathscr{N}(\widetilde{\theta}; \mu(x_{1:T}),\textup{diag}(\sigma^{2}(x_{1:T}))) \\
&\theta \sim \textup{softmax}(\widetilde{\theta}) \\
&\alpha_{t}^{(k)} = \textup{softmax}(v_{\theta}^{T}\sigma(h_{t-1}W_{h\alpha} + (\theta^{(k)}M_{global}^{(k)})W_{\theta\alpha})) \label{eq:HCRNN1_alpha} \\
&\widetilde{c}_{t} = \sum_{k=1}^{K} \alpha_{t}^{(k)}M_{global}^{(k)} \label{eq:HCRNN1_c_candiate}\\
&G_{t}^{(c)} = \sigma_{l}(x_{t}W_{xl}+h_{t-1}W_{hl}+c_{t-1}W_{cl}+b_{l}) \label{eq:HCRNN1_local_gate} \\
&c_{t} = (1-G_{t}^{(c)}) \odot c_{t-1} + G_{t}^{(c)} \odot \widetilde{c}_{t} \label{eq:HCRNN1_c}\\
&z_{t} = \sigma_{z}(x_{t} W_{xz}+h_{t-1}W_{hz} + c_{t} W_{cz} + b_{z}) \\
&r_{t} = \sigma_{r}(x_{t} W_{xr}+h_{t-1}W_{hr} + c_{t} W_{cr} + b_{r}) \label{eq:HCRNN1_reset} \\
&\widetilde{h}_{t} = (r_{t} \odot h_{t-1})W_{hh} + x_{t}W_{xh} + b_{h} \label{eq:HCRNN1_temporary_context_candidate} \\
&h_{t} = (1-z_{t}) \odot h_{t-1} + z_{t} \odot \sigma_{h}(\widetilde{h}_{t})
\end{flalign}
}
$\mu$ and $\sigma^{2}$ for the normal distribution denote the output of inference network as defined in Eq. \ref{eq:inference_mu}, \ref{eq:inference_logsigma}. $\theta$ denotes the global context proportion and its dimension is $|K|$. $M_{global}$ denotes the global context memory with $|K| \times |D|$ size. $x_t$ and $c_t$ denote the current item embedding and the local context vector respectively and they are $|D|$-dimensinoal vector. $h_t$ denote the temporary context vector with dimension $|H|$. Table \ref{table:Notation} denotes the notation for HCRNN.
\subsubsection{HCRNN-2}
After we suggest HCRNN-1, we update the generation of temporary contexts, $h_{t}$, by modifying the reset gate with the local context, $c_t$; the temporary context, $h_{t}$; and the input, $x_{t}$. Under the interest drift assumption, the interest drift can be identified if the local context and the current input are very different. If the interest drift occurred, we need to further update the temporary context, $h_{t}$, by reducing the information from $h_{t-1}$, than the case without the drift. Thus, the comparison between the local context and the current input is necessary to gauge the necessity of $h_{t-1}$. Finally, we substitute Eq. \ref{eq:HCRNN1_reset} with Eq. \ref{eq:HCRNN2_reset} as the below. 
\begin{flalign}
r_{t} &= \sigma_{r}(x_{t} W_{xr}+h_{t-1}W_{hr} + (x_{t} \odot {c_{t}}) W_{d} + b_{r}) \nonumber \\ 
 &s.t. W_{d} \geq 0 \label{eq:HCRNN2_reset}
\end{flalign}
Eq. \ref{eq:HCRNN2_reset} design the reset gate, so that the $r_t$ becomes small as the element-wise product between $c_{t}$ and $x_{t}$ decreases because of interest drift. This similarity magnitude needs to be scaled and regularized, so we multiply a constraint $W_{d} \geq 0$. Also, we used a projection operator \cite{Rakhlin2011} to handle the constraint.

\subsubsection{HCRNN-3}
The reset gate in HCRNN-2 reflects the interest drift assumption on updating the temporary context. This update in HCRNN-2 requires $x_t$, $h_{t-1}$, and $c_t$ to be mixed to generate the signal of the reset gate. The suggested update linearly models the relevance between the local context, $c_t$, and the current input, $x_t$. However, the linear activation from the element-wise product may not embody the binary nature of the temporary drift. Hence, we add a sigmoid activation on top of the element-wise product, which eventually becomes an independent gate, $G^{(d)}_t$, to model the interest drift.

Since the sigmoid function outputs a value between 0 and 1, the reset gate of HCRNN-2 in Eq. \ref{eq:HCRNN2_reset} can have a value between 0 and 1 theoretically. However, the sigmoid function is not sharp, and it makes the most LSTM forget gate values (similar to reset gate in GRU) are experimentally located in the middle state (0.5) \cite{li2018towards}. In fact, in our experiments, the reset gate, $r_t$ in Eq. \ref{eq:HCRNN2_reset} is on the average 0.47 ($\pm$ 0.03) on the CiteULike dataset. To make the temporary context focus on $x_{t}$, the value of gate multiplied by $h_{t-1}$ in Eq. \ref{eq:HCRNN1_temporary_context_candidate} need to be smaller. We model the new interest drift gate (Eq. \ref{eq:HCRNN3_drift}) and use the product of Eq. \ref{eq:HCRNN3_drift} and \ref{eq:HCRNN3_reset} in Eq. \ref{eq:HCRNN3_temporary_context_candidate}. The product of Eq. \ref{eq:HCRNN3_drift} and \ref{eq:HCRNN3_reset} has a value of 0.29 ($\pm$ 0.021) on average, and it is 38.2\% smaller than that of Eq. \ref{eq:HCRNN2_reset}.

\begin{flalign}
&G_{t}^{(d)} = \sigma_{d}((x_{t} \odot {c_{t}})W_{d}+b_{d}) \quad  s.t. W_{d} \geq 0 \label{eq:HCRNN3_drift}\\
&r_{t} = \sigma_{r}(x_{t} W_{xr}+h_{t-1}W_{hr} + b_{r}) \label{eq:HCRNN3_reset}\\
&\widetilde{h}_{t} = (r_{t} \odot (G_{t}^{(d)} \odot h_{t-1}))W_{hh} + x_{t}W_{xh} + b_{h} \label{eq:HCRNN3_temporary_context_candidate}
\end{flalign}
Eq. \ref{eq:HCRNN3_drift}, \ref{eq:HCRNN3_reset} lets the temporary context be more affected by the current input when the temporary drift is captured by the drift gate,  $G^{(d)}_t$. 

\subsection{Bi-Channel Attention and Prediction}
As mentioned in the \textit{Introduction}, it is important to learn from both long-term dependency and recent interest in a sequential recommendation. One common technique to emphasize the long-term dependency is an attention mechanism, but we introduce modified attention given a HCRNN cell structure because of its hierarchical contexts. To exploit the hierarchical contexts of HCRNN, we implement the complementary bi-channel attention as the local context attention, $\alpha^{(c)}_t$; and the temporary context attention $\alpha^{(h)}_t$. 

Both $\alpha^{(c)}_t$ and $\alpha^{(h)}_t$ needs to result in a higher attention weight if two compared context vectors are similar. $\alpha^{(h)}_t$ is modeled as a conventional linear sum based alignment function, so the training on the weight parameter can select which to attend in the temporary context. 
The projection matrices for $\alpha^{(h)}_t$ are $W_{h\alpha}^{(1)},W_{h\alpha}^{(2)}$ which are both $|H|\times|H|$ matrix.
Besides, we implemented the scaled dot-product based attention function \cite{Vaswani2017} for $\alpha^{(c)}_t$ because the dot-product will maximize the attention with the same local context vectors. This modeling will produce a stronger attention weight to the items in the similar sub-sequence. The projection matrices for $\alpha^{(c)}_t$ are $W_{c\alpha}^{(1)},W_{c\alpha}^{(2)}$ which are both $|D|\times|H|$ matrix.

We used a concatenation of $h_{t}$, $h_{t}^{(c)}$, and $h_{t}^{(h)}$ for the appropriate item prediction with a bi-linear decoding scheme following NARM as in Eq. \ref{eq:y_hat}. $W_{emb}$ is an item embedding, and $W_{B}$ is a weight for bi-linear decoding. We calculated the prediction-related loss through cross-entropy.
\begin{flalign}
&\alpha_{tj}^{(c)} = \textup{softmax}(\frac{(c_{t}W_{c\alpha}^{(1)})(c_{j}W_{c\alpha}^{(2)})^{T}}{\sqrt{|H|}}) \label{eq:att_local}\\
&\alpha_{tj}^{(h)} = \textup{softmax}(v_{h}^{T}\sigma(h_{t}W_{h\alpha}^{(1)} + h_{j}W_{h\alpha}^{(2)})) \label{eq:att_temporary} \\
&h_{t}^{(c)} = \sum_{j}\alpha_{tj}^{(c)}h_{j} \quad \textup{and} \quad h_{t}^{(h)} = \sum_{j}\alpha_{tj}^{(h)}h_{j} \\
&\widehat{y}_{t} = \textup{softmax}(W_{emb}^TW_{B}[h_{t},h_{t}^{(c)},h_{t}^{(h)}]) \label{eq:y_hat}
\end{flalign}

\subsection{Model Inference}
While training the local and the temporary contexts relies on the gradient method with a deterministic learning, HCRNN includes the global context which follows the topic probabilistic model, such as LDA, VAE, and GSM  \cite{Kingma,Miao2017}.  This generative modeling requires a maximization on the log-marginal likelihood of Eq. \ref{eq:inference_logmarginal}, so we utilize the variational inference by optimizing the evidence lower bound (ELBO) of Eq. \ref{eq:inference_elbo} \cite{Jordan1999}. 
{\small
\begin{flalign}
&\log{p(y_{1:T}|c_{1:T},h_{1:T})}=\log{\int{p(\widetilde{\theta})\prod_{t=1}p(y_{t}|\widetilde{\theta},c_{t},h_{t})}}d\widetilde{\theta} \label{eq:inference_logmarginal} \\
&\geq \sum_{t=1}^{T}E_{q(\widetilde{\theta})}[\log{p(y_{t}|\widetilde{\theta},c_{t},h_{t})}] - \operatorname{KL}[(q(\widetilde{\theta})||p(\widetilde{\theta}))] \label{eq:inference_elbo}
\end{flalign}
}%
The variational inference of HCRNN assumes the variational distribution, $q$, that is a feed-forward neural network. Following the VAE framework, $q$ also becomes the amortized inference network with the input, $x_{1:T}$, to predict $\mu$ and $\log\sigma$. Specifically, the prediction of $\mu$ is done by Eq. \ref{eq:inference_mu}, and $\log\sigma$ by Eq. \ref{eq:inference_logsigma}, where $f$ is a feed-forward neural network.

\begin{flalign}
&q(\widetilde{\theta}) = \mathscr{N}(z; \mu(x_{1:T}),\textup{diag}(\sigma^{2}(x_{1:T}))) \\
&\mu(x_{1:T}) =W_{q}^{(1)}f(x_{1:T}) + b_{q}^{(1)} \label{eq:inference_mu} \\
&\log{\sigma(x_{1:T})} =W_{q}^{(2)}f(x_{1:T}) + b_{q}^{(2)} \label{eq:inference_logsigma}
\end{flalign}
After the inference on $\mu$ and $\log\sigma$, the sampled $\widetilde{\theta}$ is used as the global context after turning it into $\theta$ by the softmax function. Our HCRNN source code is available at \url{https://github.com/gtshs2/HCRNN}.

\section{Experimental Result}
\subsubsection{Datasets} For the performance evaluation, we used three publicly available datasets: CiteULike, LastFM, and MovieLens\footnote{We converted it into a binary implicit rating by activating only the maximum rating.}. We aim at modeling a long user history, so we removed sequences whose length is less than 10. Besides, we removed the items that exist only in the test set, and the items that appeared less than 50/50/25 times in three datasets respectively. We performed cross-validation by assigning 10\% of the randomly chosen train set as the validation set. We also followed the data augmentation method as proposed in NARM \cite{li2017neural} and improved GRU4REC \cite{tan2016improved}. The data augmentation techniques can enhance the performance by reducing the overfitting. Table \ref{table:data_stat} summarizes the descriptive statistics of preprocessed datasets.
\begin{center}
    \begin{table}[ht]
        {\small
            \hfill{}
            \begin{tabular}{|c|c|c|c|}
                \hline
                \textbf{Dataset} & \textbf{CiteULike} & \textbf{LastFM} & \textbf{MovieLens} \\
                \hline
                \# sequence(train)&38,724&73,420&136,233 \\
                \hline
                \# sequence(test)&9,140&17,829&34,682 \\
                \hline
                \# clicks&1,163,813&4,575,159&5,041,882 \\
                \hline
                \# items&1,980&5,778&930 \\
                \hline
                avg. len&24.31&50.14&29.50 \\
                \hline
        \end{tabular}}
        \hfill{}
        \caption{Statistics of evaluation datasets.}
        \label{table:data_stat}
    \end{table}
\end{center}
\subsubsection{Baselines}
We compared HCRNN with the below eight baselines.
\begin{itemize}
    \item \textbf{POP} exploits the frequency of items in the training set. It always recommends items that appear most often in the training set.
    \item \textbf{SPOP} is Similar to POP, S-POP also exploits the frequency, but it recommends items that appear most often in the current sequence.
    \item \textbf{Item-KNN} \cite{Davidson2010,linden2003amazon} recommends items based on the co-occurrence number of item pairs, and Item-KNN interprets the co-occurrence as a similarity. The model recommends similar items only in the same sequence.
    \item \textbf{BPR-MF} \cite{Rendle2009} is a model representing a group of models with matrix factorization (MF) and Bayesian personalized ranking loss (BPR). By introducing the ranking loss, BPR-MF shows a better performance than a typical MF in the recommendation.
    \item \textbf{GRU4REC} \cite{hidasi2015session} is a sequential model with GRUs for the recommendation. This model adopts a session parallel batch and a loss function such as Cross-Entropy, TOP1, and BPR. 
    \item \textbf{LSTM4REC} is our version of a GRU4REC variant with LSTM.
    \item \textbf{NARM} \cite{li2017neural} is a model based on GRU4REC with an attention to consider the long-term dependency. Besides, it adopts an efficient bi-linear loss function to improve the performance with fewer parameters.
    \item \textbf{STAMP} \cite{liu2018stamp} considers both current interest and general interest of users. In particular, STAMP used an additional neural network for the current input only to model the user's current interest. Also, it proposes a tri-linear loss function. 
\end{itemize}
\subsubsection{Experiment Settings}For fair performance comparisons, we set the batch size (512), the item embedding (100), the RNN hidden dimension (100), the input dropout (0.25), the output layer dropout (0.5), the optimizer (Adam), and the learning rate (0.001)\footnote{For STAMP, we set it to 0.005 as shown in the STAMP paper.} as shown in NARM \cite{li2017neural}.
\begin{center}
    \begin{table*}[ht]
        {\small
            \hfill{}
            \begin{tabular}{|c|r|r|r|r|r|r|r|r|r|r|r|r|}
                \hline
                \textbf{}& \multicolumn{4}{c|}{\textbf{CiteULike}} & \multicolumn{4}{c|}{\textbf{LastFM}} & \multicolumn{4}{c|}{\textbf{MovieLens}}\\
                \hline
                &\multicolumn{1}{c|}{R@3}&\multicolumn{1}{c|}{R@20}&\multicolumn{1}{c|}{M@3}&\multicolumn{1}{c|}{M@20}&\multicolumn{1}{c|}{R@3}&\multicolumn{1}{c|}{R@20}&\multicolumn{1}{c|}{M@3}&\multicolumn{1}{c|}{M@20}&\multicolumn{1}{c|}{R@3}&\multicolumn{1}{c|}{R@20}&\multicolumn{1}{c|}{M@3}&\multicolumn{1}{c|}{M@20}
                \\ \hline
                POP&1.44&5.78&0.92&1.44&0.37&1.99&0.34&0.51&2.43&12.51&1.54&2.65
                \\ \hline
                S-POP&1.26&4.99&0.79&1.23&0.87&3.65&0.55&0.87&2.27&12.23&1.42&2.52
                \\ \hline
                Item-KNN&0.00&6.90&0.00&4.79&0.00&11.59&0.00&8.00&0.00&6.32&0.00&4.28
                \\ \hline
                BPR-MF&0.49&3.15&0.27&0.60&0.82&2.15&0.59&0.73&1.69&8.93&1.07&1.91
                \\ \hline
                LSTM4REC&7.07&23.33&4.93&6.82&15.29&24.75&12.68&13.95&8.52&32.80&5.63&8.45
                \\ \hline
                GRU4REC&\setul{0.5pt}{.4pt} \ul{8.37}&24.19&\setul{0.5pt}{.4pt} \ul{5.98}&\setul{0.5pt}{.4pt} \ul{7.86}&18.29&26.46&\setul{0.5pt}{.4pt} \ul{15.85}&\setul{0.5pt}{.4pt} \ul{16.95}&8.50&32.74&5.60&8.42
                \\ \hline
                NARM&7.81&\setul{0.5pt}{.4pt} \ul{24.82}&5.40&7.41&\setul{0.5pt}{.4pt} \ul{18.30}&\setul{0.5pt}{.4pt} \ul{33.60}&13.12&15.25&\setul{0.5pt}{.4pt} \ul{9.14}&\setul{0.5pt}{.4pt} \ul{33.42}&\setul{0.5pt}{.4pt} \ul{6.09}&\setul{0.5pt}{.4pt} \ul{8.93}
                \\ \hline
                STAMP&5.09&21.93&3.25&5.22&9.29&19.84&6.62&8.01&3.95&20.52&2.65&4.47
                \\ \hline \hline 
                HCRNN- 1&8.60&25.36&6.18&8.16&20.67\textsuperscript{*}&34.40\textsuperscript{*}&15.77&17.68\textsuperscript{*}&9.23&33.78\textsuperscript{*}&6.13&9.00
                \\ \hline
                HCRNN- 2&8.83&25.10&6.41\textsuperscript{*}&8.38\textsuperscript{*}&20.78\textsuperscript{*}&34.14\textsuperscript{*}&16.20&18.08\textsuperscript{*}&9.22&33.76\textsuperscript{*}&6.14&9.01
                \\ \hline
                HCRNN- 3&9.21\textsuperscript{*}&25.42\textsuperscript{*}&6.65\textsuperscript{*}&8.61\textsuperscript{*}&21.39\textsuperscript{*}&34.72\textsuperscript{*}&16.66\textsuperscript{*}&18.52\textsuperscript{*}&9.38\textsuperscript{*}&33.67\textsuperscript{*}&6.23\textsuperscript{*}&9.08\textsuperscript{*}
                \\ \hline
                HCRNN-3 + Bi&\textbf{9.33\textsuperscript{*}}&\textbf{25.81\textsuperscript{*}}&\textbf{6.74\textsuperscript{*}}&\textbf{8.70\textsuperscript{*}}&\textbf{21.90\textsuperscript{*}}&\textbf{34.80\textsuperscript{*}}&\textbf{17.33\textsuperscript{*}}&\textbf{19.12\textsuperscript{*}}&\textbf{9.53\textsuperscript{*}}&\textbf{33.83\textsuperscript{*}}&\textbf{6.38\textsuperscript{*}}&\textbf{9.21\textsuperscript{*}}
                \\
                \hline \hline
                Improvement(\%)&11.47&3.99&12.71&10.69&19.67&3.57&9.34&12.80&4.27&1.23&4.76&3.14
                \\ \hline            
        \end{tabular}}
        \caption{Performance evaluation of the proposed models. The boldface indicates the best result among our models and the underline indicates the best result among the baselines. $P^{*}<0.05$ (Student's $t$-test)} 
        \label{table:quant_results}
    \end{table*}
\end{center}
\subsection{Quantitative Performance Evaluation}
Table \ref{table:quant_results} shows the performance of the baselines and HCRNN with two measurements of \textit{recall} at K (R@K) and \textit{mean reciprocal ranking} at K (M@K), which are widely used in the sequential recommendation. We varied K by 3 and 20. The experiments on HCRNN has an ablation study variation of HCRNN-1, HCRNN-2, HCRNN-3, and HCRNN-3 with bi-channel attentions (HCRNN-3+Bi). The quantitative evaluation indicates that the variations of HCRNN have significant performance improvements in all data and metrics. Particularly, HCRNN-3 with the bi-channel attentions always exhibits the best performance. 
Additionally, the better performance of HCRNN over NARM, which also has a context modeling, may suggest the need for hierarchical context modeling in recommendations. Moreover, HCRNN shows the best result compared to the RNN based recommendations, i.e., NARM, GRU4REC, and LSTM4REC, so the modified HCRNN cell may have contributed to the performance improvements. 
As HCRNN-3 with drift gate, $G_{t}^{(d)}$, shows better results than HCRNN-1 and HCRNN-2, our interest drift assumption may be experimentally justifiable. As HCRNN-3+Bi is the best case, we justify that bi-channel attention with hierarchical contexts may improve the performance experimentally.
Finally, NARM with RNN and attention shows better performance than STAMP with a feed-forward neural network. This demonstrates the importance of sequential modeling in recommendations with long sequences.

\subsection{Qualitative Analysis}
From the sensitivity perspective, the global context is unlikely to change given a single item. The local context should change when the item selection is dissimilar to the previous selection, but if the selections are similar, the local context does not change much by Eq. \ref{eq:HCRNN1_alpha}-\ref{eq:HCRNN1_c}. The temporary context likely changes for each selected item to represent the current interest and the temporary context significantly changes when the genre transition happens. Because of the two gating structure of Eq. \ref{eq:HCRNN3_temporary_context_candidate} modeling, the average amount of change in the temporary context is more significant than that of the local context as our assumption and expectation. Experimentally, with CiteULike, the temporary context, h, changes 0.278($\pm$ 0.037) on average at every timestep, and the local context, c, changes by 0.005($\pm$ 0.024) on average.

\subsubsection{Context Embedding}
\begin{figure}[ht]
    \includegraphics[width=1.0\columnwidth]{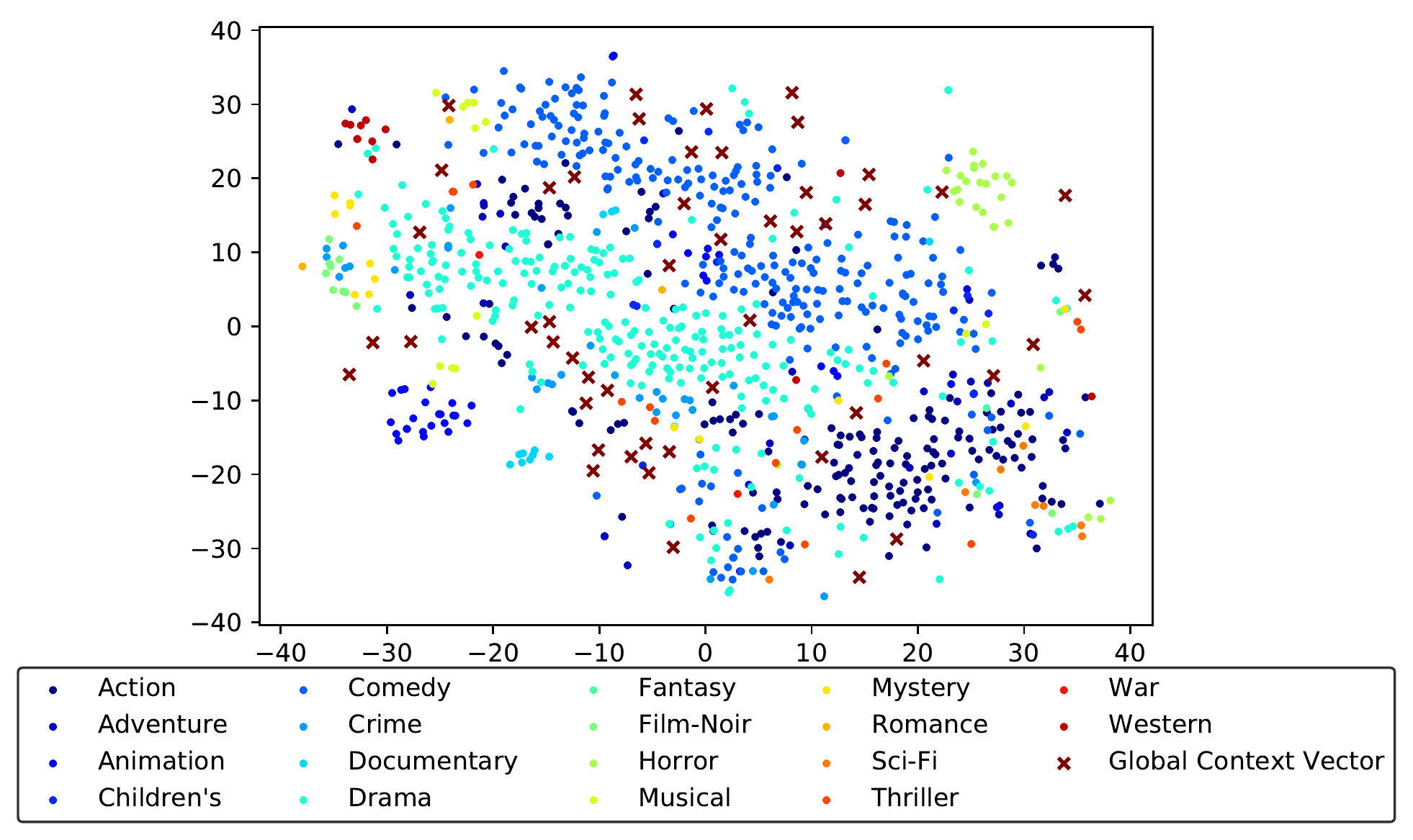}
    \caption{Item embedding and global context vector, $M_{global}^{(k)}$, visualization with tSNE\cite{Maaten2008}. Item embedding is interpretable with genre, and global context vector cover the most of the items.}
    \label{fig:embedding_tsne}
\end{figure}

This study models the hierarchical context, global context memory ($M_{global}$), local context ($c_{t}$), and the temporary context ($h_{t}$). 
The local context is generated by the global context memory, and the temporary context is generated by the previous temporary context and the current item embedding ($x_{t}$).
The first analysis is visualizing the global context memory ($M_{global}$) and the item embedding ($x_{t}$), to verify the quality of inputs to the construction on the local context ($c_{t}$) and the temporary context ($h_t$). Figure \ref{fig:embedding_tsne} is the joint visualization of the item and the global context memory. The item embeddings are coherently organized as a cohesive cluster with the same genre, and the global context memory covers most of the area that the item embeddings are dispersed. 
Given the item and the global context memory, we calculated the cosine similarity, and Table \ref{table:context_embedding} enumerates the most aligned items with a specific global context vector in the global context memory.

\begin{center}
    \begin{table}[ht]
        {\small
            \hfill{}
            \begin{tabular}{|c|c|c|}
                \hline
                &\textbf{Genre} & \textbf{Movie Title} \\
                \hline \hline
                \multirow{ 2}{*}{$M_{global}^{(6)}$}&\multirow{ 2}{*}{Animation} &Pinocchio, Yellow Submarine, \\
                &&Snow White and the Seven Dwarfs \\
                \hline
                \multirow{ 2}{*}{ $M_{global}^{(19)}$}&\multirow{2}{*}{Action}&Star Trek: Generations, Predator, \\
                &&Butch Cassidy and the Sundance Kid \\
                \hline
                \multirow{ 2}{*}{$M_{global}^{(31)}$}&\multirow{2}{*}{Horror} &Scream, An American Werewolf \\
                &&in London, Dracula \\
                \hline
        \end{tabular}}
        \hfill{}
        \caption{Interpretation of global context vector. We listed the items (movie title) close to each global context vector.}
        \label{table:context_embedding}
    \end{table}
\end{center}
\subsubsection{Gate Analysis}
\begin{figure}[t!]
    \centering
    \begin{subfigure}[t]{0.23\textwidth}
        \centering
        \includegraphics[width=\textwidth]{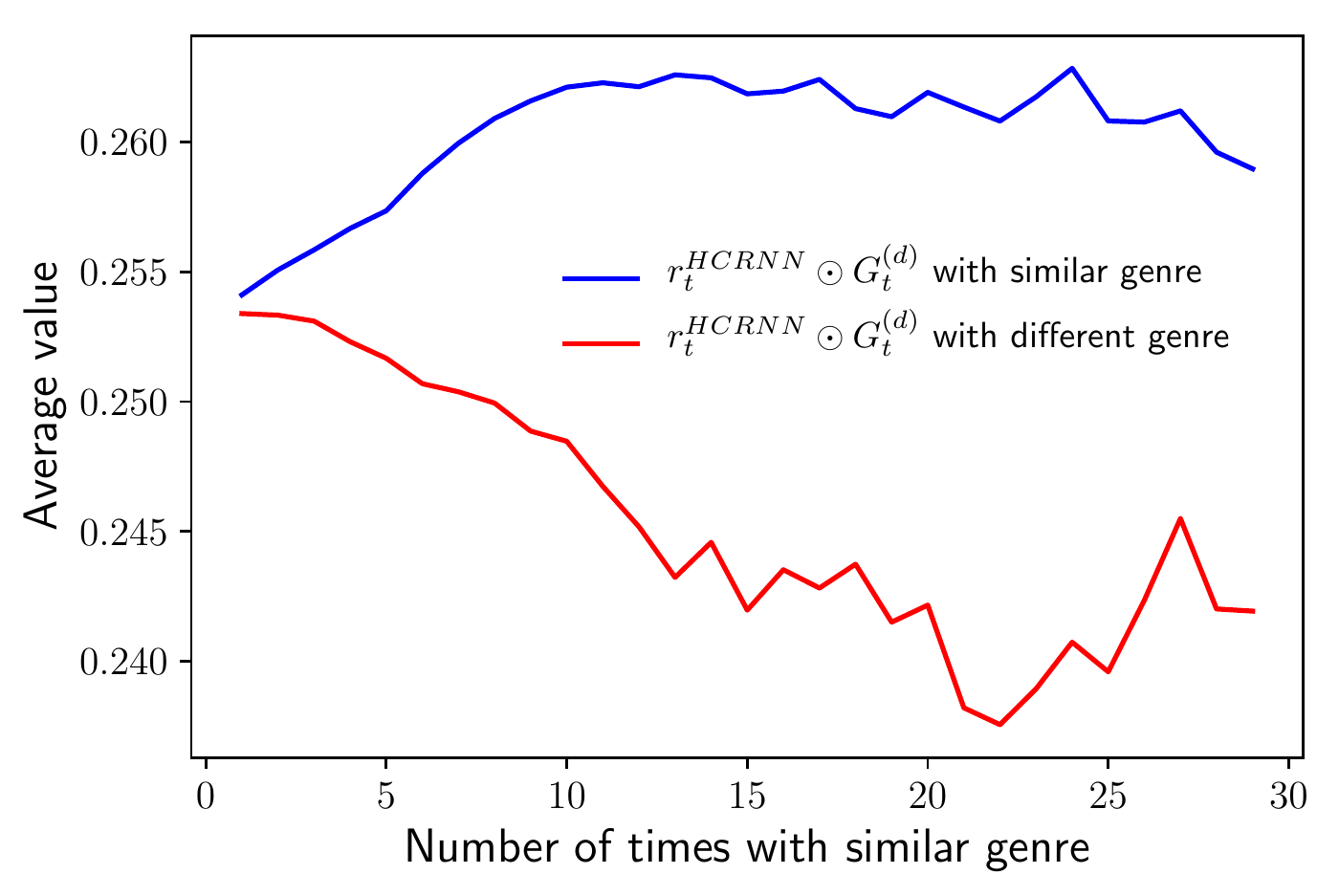}
        \caption{Average value of $r^{HCRNN}_{t} \odot G_{t}^{(d)}$  gate after appearing items with similar genre consecutively.}
        \label{fig:rT2_ma_5}
    \end{subfigure}
    \hfill
    \begin{subfigure}[t]{0.23\textwidth}
        \centering
        \includegraphics[width=\textwidth,height=2.8cm]{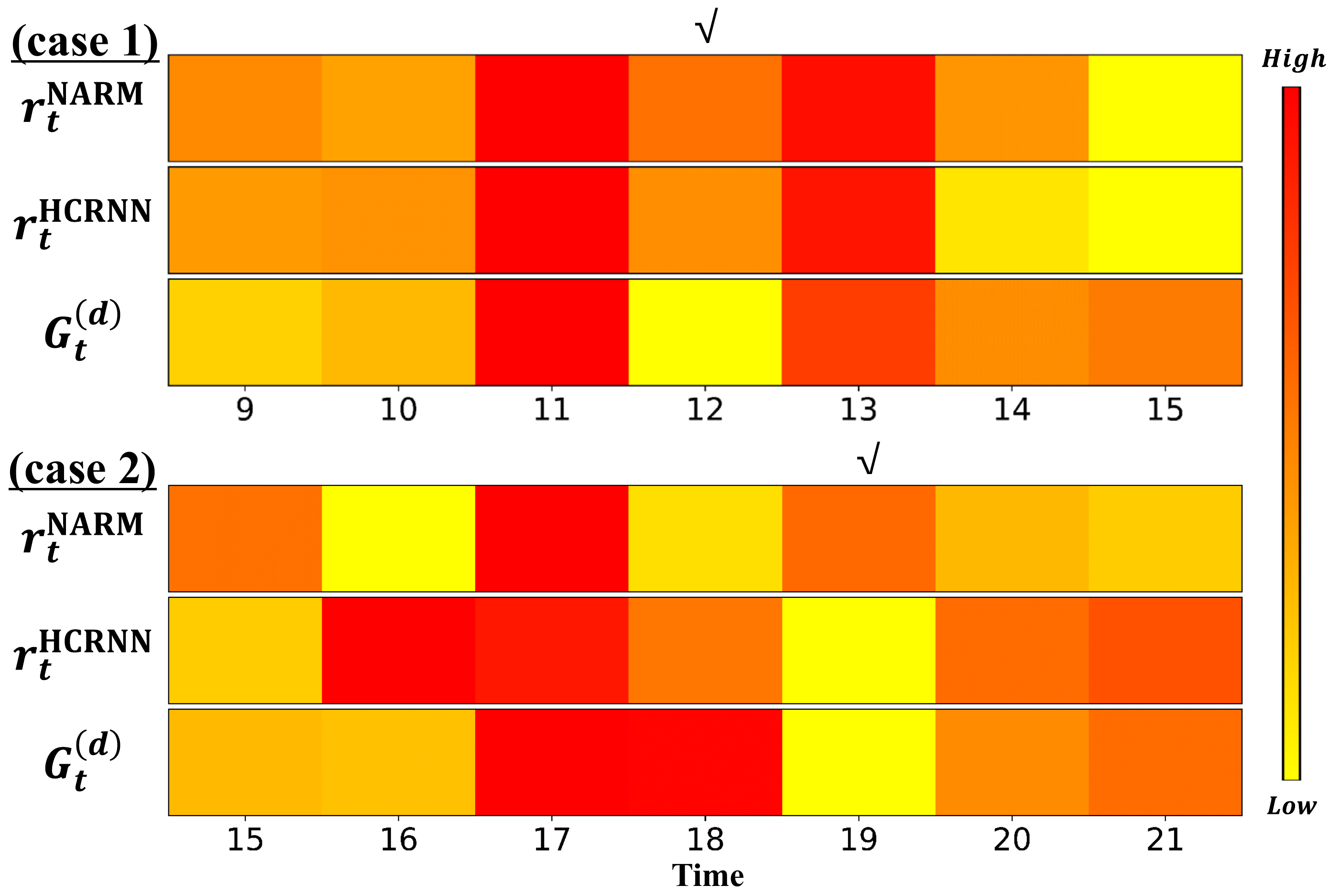}
        \caption{Gate heatmap for a user history as an example. We mark ``check" when the genre of item changes.}
        \label{fig:gate_heatmap_case12}
    \end{subfigure}
    \caption{HCRNN takes a large value of $r^{HCRNN}_{t} \odot G_{t}^{(d)}$ if the current input of item has similar genre with previous input of item. In the opposite case, HCRNN grasps the user's interest drift and changes $r^{HCRNN}_{t} \odot G_{t}^{(d)}$ to smaller.}
    \label{fig:gate_analysis}
\end{figure}

HCRNN is a model to capture the user's interest drift with hierarchical context and drift gate $G^{(d)}_{t}$. For the comparison of HCRNN and NARM gate structures, we define $r_{t}^{HCRNN}$ and $r_{t}^{NARM}$ as the HCRNN and NARM reset gates, respectively.
In order to incorporate the interest drift assumption, we designed the value of $r_{t}^{HCRNN} \odot G^{(d)}_{t}$ gate applied to $h_{t-1}$ to be smaller when updating $\widetilde{h}_{t}$ in Eq. \ref{eq:HCRNN3_temporary_context_candidate} if interest drift occurred.
Figure \ref{fig:gate_analysis} represents the gate value when the user clicks the same genre of an item as the previous step and when the user does not.
The x-axis in Figure \ref{fig:rT2_ma_5} represents the number of consecutive items which has the same genre until the right before timestep.
In general, if the genre of the current input is different with previous items, $r_{t}^{HCRNN} \odot G^{(d)}_{t}$ has a smaller value compared to the opposite situation.
Besides, when a user clicked the same genre of items consecutively, and instantly clicks a different genre of items, the value of $r_{t}^{HCRNN} \odot G^{(d)}_{t}$ becomes smaller.
\subsubsection{Bi-Channel Attention} 
\begin{figure}[t!]
    \centering
    \begin{subfigure}[t]{0.23\textwidth}
        \centering
        \includegraphics[width=\textwidth]{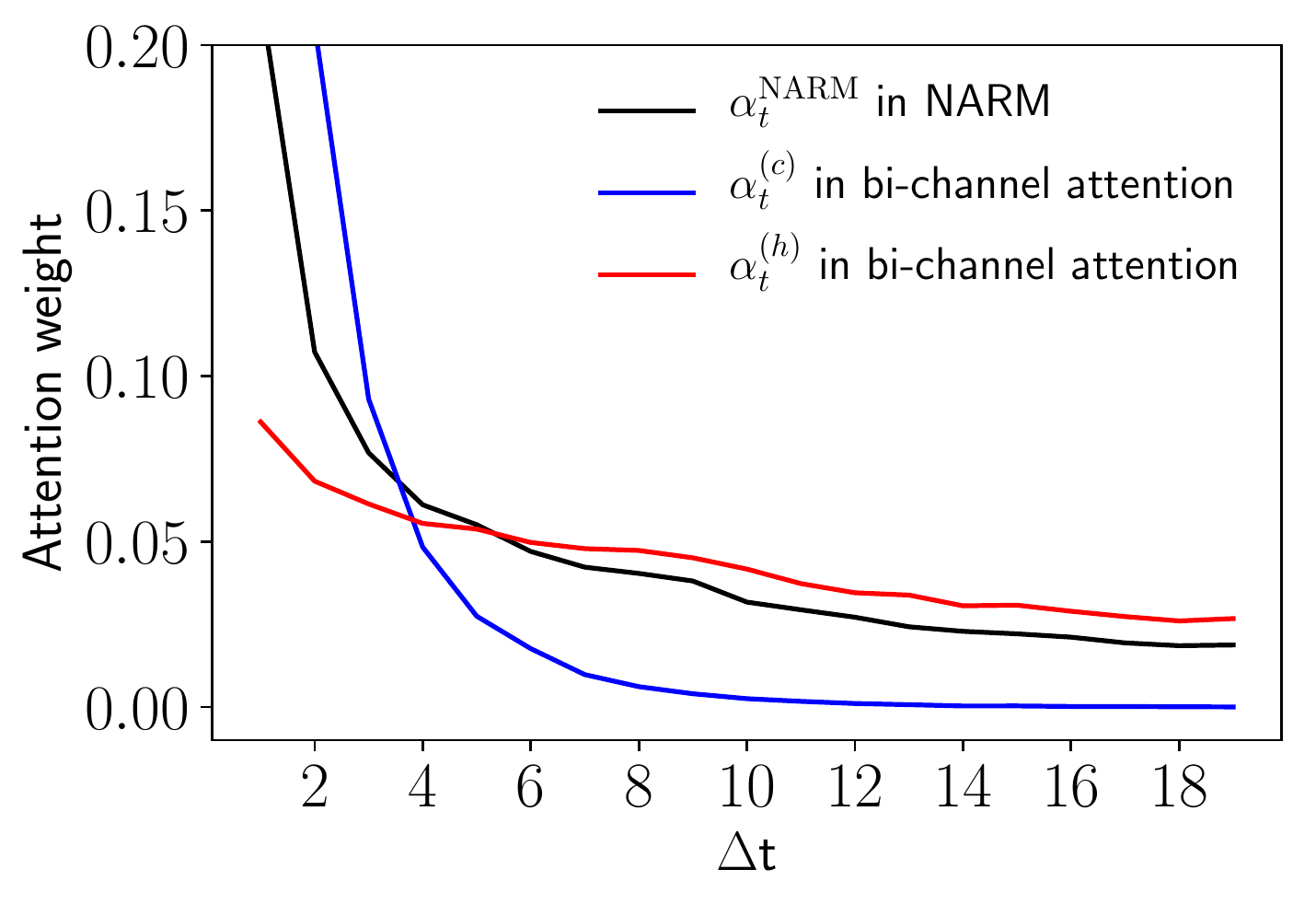}
        \caption{Averaged attention weight over time difference. $\Delta{t}$ means a time difference between prediction time step and the timestep of the previous user history.}
        \label{fig:qualitative_attention}
    \end{subfigure}
    \hfill
    \begin{subfigure}[t]{0.23\textwidth}
        \centering
        \includegraphics[width=\textwidth,height=2.8cm]{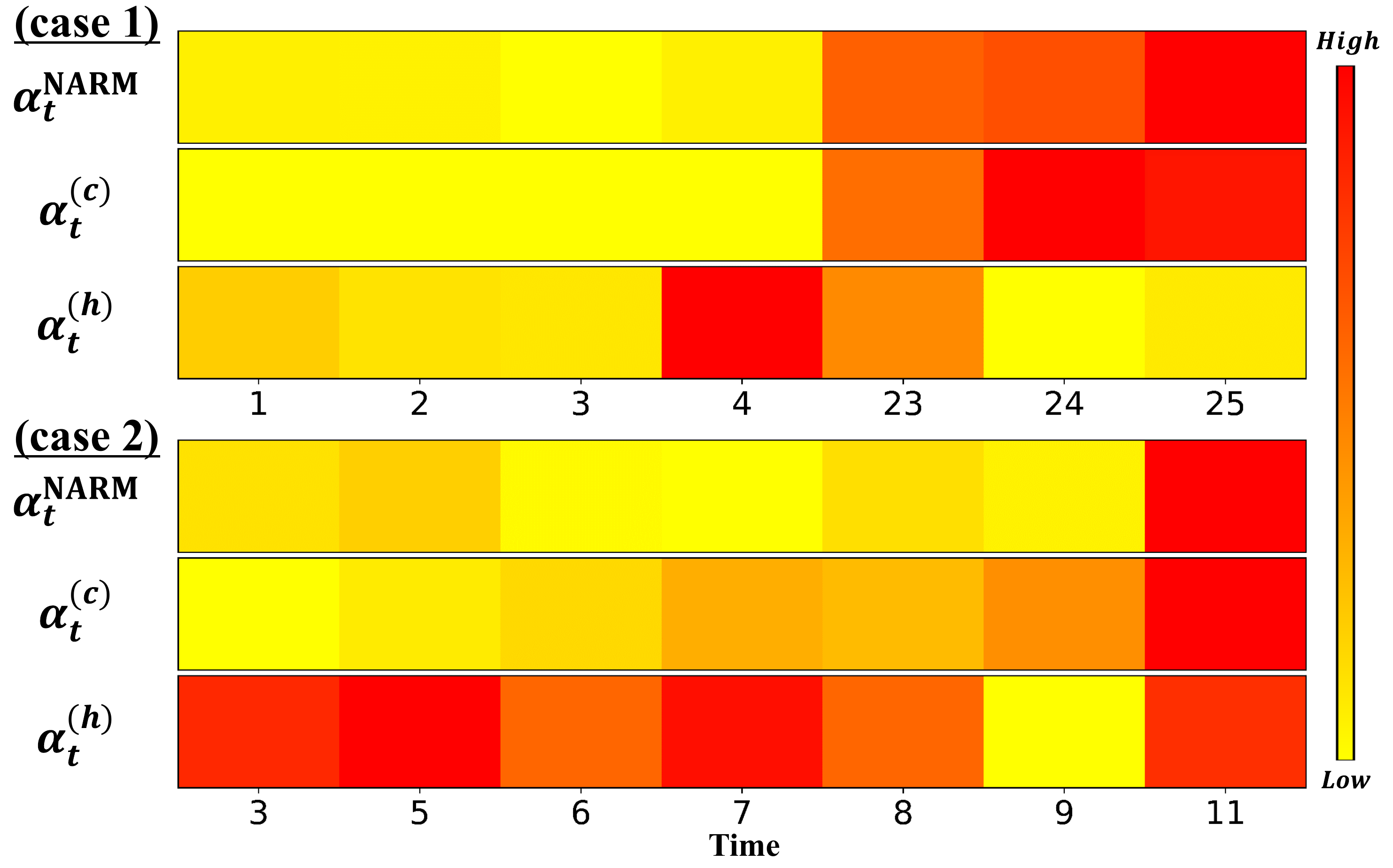}
        \caption{Attention heatmap for a user history. The first row of each case is an attention weight in NARM, and the two below are our bi-channel attention weights.}
        \label{fig:HCRNN_attention_example}
    \end{subfigure}
    \caption{Temporary context based attention, $\alpha_{t}^{(h)}$, in HCRNN is spread over a long period relatively. Local context based attention, $\alpha_{t}^{(c)}$, in HCRNN has a large value on recent user records.}
    \label{fig:qualitative_attention_overall}
\end{figure}

As mentioned in section \textit{Introduction}, it is important to understand both long-term dependency and recent interests in recommendations with a long user history. Therefore, we present the bi-channel attentions from local and temporary contexts \textit{Bi-Channel Attention and Prediction}, and we present the result in Figure \ref{fig:qualitative_attention_overall}. Figure \ref{fig:qualitative_attention} shows the averaged attention weights over the test user histories. NARM has a single attention mechanism, so NARM attention weight, $\alpha_{t}^{NARM}$, cannot differentiate the attentions on the local and the temporary contexts. However, the bi-channel attentions distinguishes the attentions for the sub-sequence continuation, $\alpha_{t}^{(c)}$; and for the temporary transitions, $\alpha_{t}^{(h)}$. Figure \ref{fig:qualitative_attention} and \ref{fig:HCRNN_attention_example} indicates that $\alpha_{t}^{(c)}$ focuses on the neighbor attention to check the continuation; and that $\alpha_{t}^{(h)}$ spreads out through the whole sequence to check the similar temporary transition. 

\subsubsection{Case Study}
\begin{figure}[t!]
    \includegraphics[width=1.0\columnwidth]{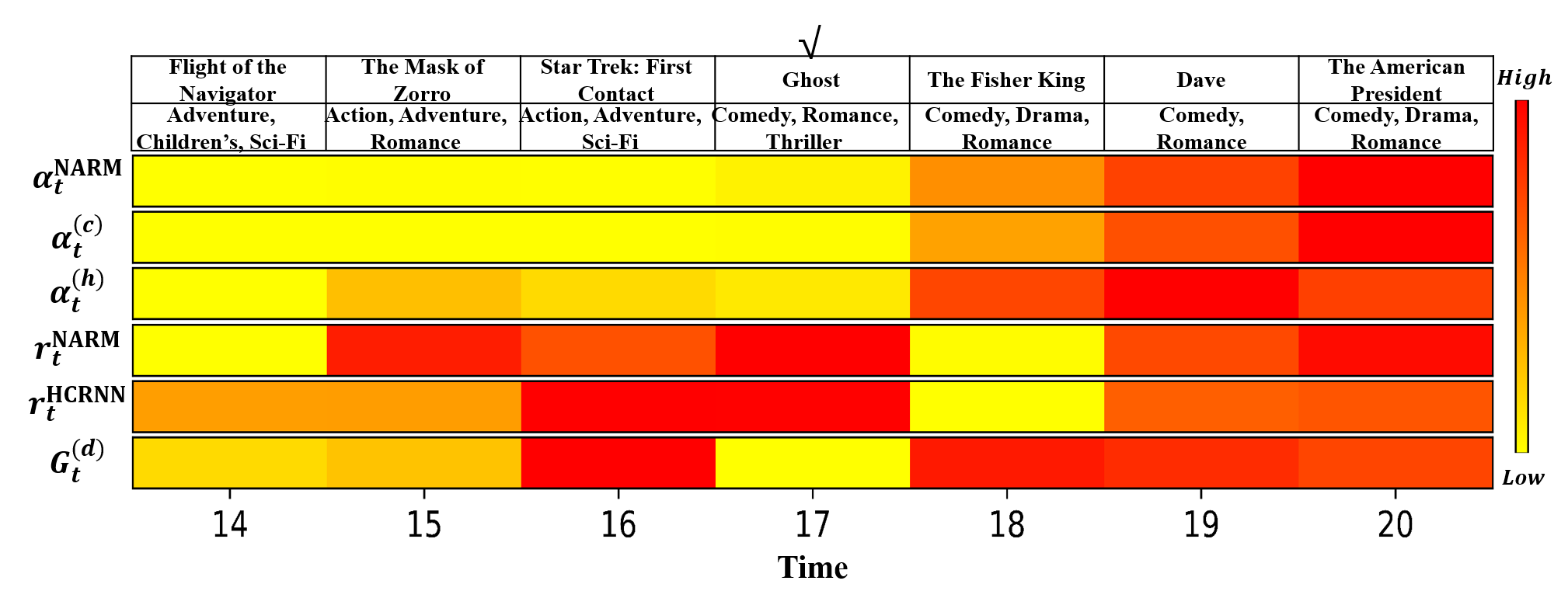}
    \caption{Attention, gate value in NARM and HCRNN, and the change of context value in HCRNN overtime. The drift gate $ G^{(d)}_{t=17} $ in HCRNN captured the temporary interest drift}
    \label{fig:drift_reset_gate_ex}
\end{figure}

Figure \ref{fig:drift_reset_gate_ex} shows the attention weights, and the gate values for selected user history. In Figure \ref{fig:drift_reset_gate_ex}, the top three rows represent the attention weights comparing NARM and HCRNN. The bi-channel attentions of HCRNN results in two rows of attentions. The attention of local contexts, $\alpha_{t}^{(c)}$, focuses on recent history, and the attention of temporary contexts, $\alpha_{t}^{(h)}$, considers relatively far history. These mean that $\alpha_{t}^{(c)}$ emphasizes belonging to the same sub-sequence, and $\alpha_{t}^{(h)}$ tries to find the similar transition throughout the entire history. In particular, $\alpha^{(h)}_{t=15}$ has a relatively high attention weight compared to $\alpha^{(c)}_{t=15}$ and $\alpha^{NARM}_{t=15}$. The rational behind this temporary high attention originates from the same genre of the item entered as input at the last timestep, $t = 20$, whose genre is Romance.

After we observe the attention weights, we observe the gate values, which are $r^{NARM}_t$, $G^{(d)}_t$, and $r^{HCRNN}_t$, to verify that the gate operates as we expected. We observed that $G^{(d)}_{t=17}$ has a relatively small value. This small value is caused by the selection of items disaligned to the previous sub-sequence at $t=16$. This phenomenon demonstrates that $G^{(d)}_t$ can capture the interest drift to reset the temporary context to accept further information from the current item embedding of $x_t$, as designed in Eq. \ref{eq:HCRNN3_drift}.  As $G^{(d)}_t$ is controlled by the local context, the discontinuation of genre matters in  $G^{(d)}_t$. However, $r^{HCRNN}_t$, which also controls the reset of the temporary context in Eq. \ref{eq:HCRNN3_reset}, is not activated because $r^{HCRNN}_t$ only takes the temporary context as the inputs, so the discontinuation does not matter in $r^{HCRNN}_t$. This rationale applies to $r^{NARM}_t$, as well. 
On the contrast, the user history has the same genre at $t = 18,19,20$, so $G^{(d)}_t$ also keeps high gate values to prevent the reset on the temporary context.

\section{Conclusion}
This paper proposes HCRNN to model the hierarchical contexts for recommendations. 
We have separated the creation of temporary contexts and local contexts, and it helps the temporary context to focus on the more current item and transient interest.
For effective hierarchical context modeling, we present a new context generation structure that utilizes the advantages of the latent topic model and the memory network to contain the abstract information for the global and the local contexts. We also propose the new gate mechanism to incorporate the interest drift assumption.  
To support HCRNN with hierarchical contexts, we propose bi-channel attentions to account for both long-term dependency and recent interest in the long user history.
\subsubsection{Acknowledgments.}
This research was supported by Basic Science Research Program through the National Research Foundation of Korea(NRF) funded by the Ministry of Education(NRF-2018R1C1B6008652))
\bibliography{aaai19_HCRNN}
\bibliographystyle{aaai}
\end{document}